\documentclass[twocolumn,showpacs,preprintnumbers,amsmath,amssymb]{revtex4}
\usepackage{graphicx}
\usepackage{latexsym} 
\usepackage{amssymb}
\usepackage{textcomp}
\usepackage{bm}
\usepackage{threeparttable}

\begin{document}

\hyphenation{dis-tri-bu-tion dis-tri-bu-tions 
oli-go-nu-cleo-tide Arab-i-dop-sis 
Meth-a-no-co-ccus jann-a-schii Tre-po-nema pall-i-dum Vib-rio chol-e-rae
Haem-o-phi-lus infl-uen-zae Chla-my-dia mur-i-da-rum du-pli-cat-ion 
re-pli-cat-ion chro-mo-some Ar-chaeo-glo-bus Stre-pto-co-ccus pneu-mo-niae
Clos-tri-dium ace-to-bu-tyli-cum char-ac-terized}
\def\Atha{{\it Arabidopsis thaliana}}  \def\atha{{\it A. thaliana}}
\def\Aaeo{{\it Aquifex aeolicus}}  \def\aaeo{{\it A. aeolicus}}
\def\Aful{{\it Archaeoglobus fulgidus}}    \def\aful{{\it A. fulgidus}}  
\def\Aper{{\it Aeropyrum pernix}}  \def\aper{{\it A. pernix}} 
\def\Atum{{\it Agrobacterium tumefaciens}}  \def\atum{{\it A. tumefaciens}} 
\def\Bbur{{\it Borrelia burgdorferi}}  \def\bbur{{\it B. burgdorferi}}  
\def\Bfra{{\it Bacteroides fragilis}}  \def\bfra{{\it B. fragilis}}
\def\Bhal{{\it Bacillus halodurans}} \def\bhal{{\it B. halodurans}} 
\def\Bmel{{\it Brucella melitensis}}   \def\Bmel{{\it B. melitensis}}   
\def\Bsub{{\it Bacillus subtilis}} \def\bsub{{\it B. subtilis}} 
\def\Busp{{\it Buchnera sp. APS}}   \def\busp{{\it B. sp.}}   
\def\Baph{{\it Buchnera aphidicola}}   \def\baph{{\it B. aphidicola}}   
\def\Cele{{\it Caenorhabditis elegans}} \def\cele{{\it C. elegans}}
\def\Cjej{{\it Campylobacter jejuni}} \def\cjej{{\it C. jejuni}} 
\def\Ccre{{\it Caulobacter crescentus}} \def\ccre{{\it C. crescentus}} 
\def\Cvib{{\it Caulobacter vibrioides}}  \def\cvib{{\it C. vibrioides}} 
\def\Clim{{\it Chlorobium limicola}} \def\clim{{\it Ch. limicola}} 
\def\Cmur{{\it Chlamydia muridarum}}  \def\cmur{{\it C. muridarum}}  
\def\Cace{{\it Clostridium acetobutylicum}}  \def\cace{{\it C. acetobutylicum}}
\def\Cper{{\it Clostridium perfringens}}  \def\cper{{\it C. perfringens}}   
\def\Ctet{{\it Clostridium tetani E88}}  \def\ctet{{\it C. tetani}}   
\def\Cpne{{\it Chlamydia pneumoniae}}  \def\cpne{{\it Ch. pneumoniae}}  
\def\Ctra{{\it Chlamydia trachomatis}} \def\ctra{{\it Ch. trachomatis}} 
\def\Cglu{{\it Corynebacterium glutamicum}} \def\cglu{{\it C. glutamicum}}  
\def\chyd{{\it C. hydrogenoformans}}
\def\Drad{{\it Deinococcus radiopugans}} \def\drad{{\it D. radiopugans}} 
\def\Dmel{{\it Drosophila melanogaster}} \def\dmel{{\it D. melanogaster}}

\def\Ecol{{\it Escherichia coli}} \def\ecol{{\it E. coli}} 
\def\Ecun{{\it Encephalitozoon cuniculi}} \def\ecun{{\it E. cuniculi}} 
\def\Fhep{{\it Flavobacterium heparinum}} \def\fhep{{\it F. heparinum}} 
\def\Fnuc{{\it Fusobacterium nucleatum}} \def\fnuc{{\it F. nucleatum}} 
\def\ftul{{\it F. tularensis}}
\def\Gmax{{\it Glycine max}}  \def\gmax{{\it G. max}}    
\def\Hasp{{\it Halobacterium sp.}}  \def\hasp{{\it H. sp.}} 
\def\Haur{{\it Herpetosiphon aurantiacus}} \def\haur{{\it H. aurantiacus}} 
\def\Hinf{{\it Haemophilus influenzae}} \def\hinf{{\it H. influenzae}} 
\def\Hpyl{{\it Helicbacter pylori}}  \def\hpyl{{\it H. pylori}}  
\def\Hsap{{\it Homo sapiens}} \def\hsap{{\it H. sapiens}} 
\def\Hvol{{\it Halobacterium volcanii}}  \def\hvol{{\it H. volcanii}}  
\def\Llac{{\it Lactococcus lactis}}  \def\llac{{\it L. lactis}}   
\def\Lmon{{\it Listeria monocytogenes}} \def\lmon{{\it L. monocytogenes}} 
\def\Msuc{{\it Mannheimia succiniciproducens}} \def\msuc{{\it M. succiniciproducens}}
\def\Mlot{{\it Mesorhizobium loti}}  \def\mlot{{\it M. loti}}  
\def\Mfer{{\it Methanothermus fervidus}} \def\mfer{{\it M. fervidus}} 
\def\Mjan{{\it Methanococcus jannaschii}} \def\mjan{{\it M. jannaschii}} 
\def\Mthe{{\it Methanobacterium thermoautotrophicum}} 
                     \def\mthe{{\it M. thermoautotrophicum}}  
\def\Mmus{{\it Mus musculus}} \def\mmus{{\it M. musculus}} 
\def\Mgen{{\it Mycoplasma genitalium}} \def\mgen{{\it M. genitalium}} 
\def\Mpen{{\it Mycoplasma penetrans}} \def\mpen{{\it M. penetrans}} 
\def\Mpne{{\it Mycoplasma pneumoniae}} \def\mpne{{\it M. pneumoniae}} 
\def\Mpul{{\it Mycoplasma pulmonis}} \def\mpul{{\it M. pulmonis}} 
\def\Mlep{{\it Mycobacterium leprae}} \def\mlep{{\it M. leprae}} 
\def\Mtub{{\it Mycobacterium tuberculosis}} \def\mtub{{\it M. tuberculosis}} 
\def\Nmen{{\it Neisseria meningitidis}} \def\nmen{{\it N. meningitidis}} 
\def\Neis{{\it Neisseria}}
\def\Nost{{\it Nostoc sp.}} \def\nost{{\it N. sp.}} 
\def\Paby{{\it Pyrococcus abyssi}}  \def\paby{{\it P. abyssi}}   
\def\Paero{{\it Pyrobaculum aerophilum}}  \def\paero{{\it P. aerophilum}}   
\def\Paeru{{\it Pseudomonas aeruginosa}} \def\paeru{{\it P. aeruginosa}} 
\def\Pfur{{\it Pyrococcus furiosus}}  \def\pfur{{\it P. furiosus}}   
\def\Phor{{\it Pyrococcus horikoshii}} \def\phor{{\it P. horikoshii}} 
\def\Pfal{{\it Plasmodium falciparum}} \def\pfal{{\it P. falciparum}}
\def\Plas{{\it Plasmodium}}
\def\Pmul{{\it Pasteurella multocida}} \def\pmul{{\it P. multocida}} 
\def\Rcon{{\it Rickettsia conorii}} \def\rcon{{\it R. conorii}} 
\def\Rpro{{\it Rickettsia prowazekii}} \def\rpro{{\it R. prowazekii}} 
\def\Rsol{{\it Ralstonia solanacearum}} \def\rsol{{\it R. solanacearum}} 
\def\Saur{{\it Staphylococcus aureus}} \def\saur{{\it S. aureus}} 
\def\Sent{{\it Salmonella enterica}} \def\sent{{\it S. enterica}} 
\def\Scer{{\it Saccharomyces cerevisiae}} 
                     \def\scer{{\it S. cerevisiae}} 
\def\Smel{{\it Sinorhizobium meliloti}}  \def\smel{{\it S. meliloti}}  
\def\Smel{{\it Sinorhizobium meliloti}}  \def\smel{{\it S. meliloti}}  
\def\Sfle{{\it Shigella flexneri}} \def\sfle{{\it S. flexneri}}  
\def\Sone{{\it Shewanella oneidensis}} \def\sone{{\it S. oneidensis}}  
\def\Spne{{\it Streptococcus pneumoniae}} \def\spne{{\it S. pneumoniae}}  
\def\Spyo{{\it Streptococcus pyogenes}} \def\spyo{{\it S. pyogenes}} 
\def\Scoe{{\it Streptomyces coelicolor}} \def\scoe{{\it S. coelicolor}} 
\def\Save{{\it Streptomyces avermitilis}} \def\save{{\it S. avermitilis}} 
\def\Ssol{{\it Sulfolobus solfataricus}} \def\ssol{{\it S. solfataricus}} 
\def\Stok{{\it Sulfolobus tokodaii}} \def\stok{{\it S. tokodaii}}  
\def\Stub{{\it Solanum tuberosum}} \def\stub{{\it S. tuberosum}} 
\def\Styp{{\it Salmonella typhimurium LT2}} \def\styp{{\it S. typhimurium}} 
\def\Sent{{\it Salmonella enterica}} \def\sent{{\it S. enterica}} 
\def\Syne{{\it Synechococcus sp.}} \def\syne{{\it S. sp.}}
\def\Taci{{\it Thermoplasma acidophilum}}  \def\taci{{\it T. acidophilum}}   
\def\Tmar{{\it Thermotoga maritima}} \def\tmar{{\it T. maritima}} 
\def\Tpal{{\it Treponema pallidum}} \def\tpal{{\it T. pallidum}} 
\def\Tten{{\it Thermoprotues tenax}}  \def\tten{{\it T. tenax}}  
\def\Tvol{{\it Thermoplasma volcanium}} \def\tvol{{\it T. volcanium}}  
\def\Telo{{\it Thermosynechococcus elongatus}} \def\telo{{\it T. elongatus}}  
\def\Tthe{{\it Thermus thermophilus}}  \def\tthe{{\it T. thermophilus}}
\def\Uure{{\it Ureaplasma urealyticum}} \def\uure{{\it U. urealyticum}} 
\def\Vcho{{\it Vibrio cholerae}} \def\vcho{{\it V. cholerae}} 
\def\Xfas{{\it Xylella fastidiosa}} \def\xfas{{\it X. fastidiosa}} 
\def\Ypes{{\it Yersinia pestis}}  \def\ypes{{\it Y. pestis}}  
\def\Rnor{{\it Rattus norvegicus}}  \def\rnor{{\it R. norvegicus}}
\def\Spom{{\it Saccharomyces pombe }}  \def\spom{{\it S. pombe}} 
\def\Cfam{{\it Canis familiaris}}  \def\cfam{{\it C. familiaris}} 
\def\Ggal{{\it Gallus gallus}}  \def\ggal{{\it G. gallus}} 
\def\Btau{{\it Bos taurus}}  \def\btau{{\it B. taurus}} 
\def\Ptro{{\it Pan troglodytes}}  \def\ptro{{\it P. troglodytes}} 
\def\Amel{{\it Apis mellifera}}  \def\amel{{\it A. mellifera}} 
\def\Drer{{\it Danio rerio}}  \def\drer{{\it D. rerio}} 
\def\Agam{{\it Anophele gambiae}} \def\agam{{\it A. gambiae}} 
\def\Mmul{{\it Macaca multatta}}  \def\mmul{{\it M. multatta}} 
\def\Tcas{{\it Tribolium casteneum}}  \def\tcas{{\it T. casteneum}} 
\def\Osat{{\it Oryza sativa}} \def\osat{{\it O. sativa}} 
\def\Fcam{{\it Fritillaria camschatcensis}} \def\fcam{{\it F. camschatcensis}} 
\def\ecols{{\it E. col.}}  \def\mjans{{\it M. jan.}} 
\def\cmurs{{\it C. mur.}}  \def\tpals{{\it T.pal}}
\def\newpar{{\par\noindent}}
\def\skipaline{{\vskip 12pt plus 1pt}}
\def\skiphafline{{\vskip 6pt plus 1pt}}
\def\qline{{\vskip 3pt plus 1pt}} \def\hfline{{\skiphafline}}
\def\newpar{{\hfline\noindent}}
\def\sslash{{\slash\hskip -5pt}}
\def\slasha{{\sslash a}}
\def\mn{{\medskip\par\noindent}}
\def\bn{{\bigskip\par\noindent}}
\def\sn{{\smallskip\par\noindent}}
\def\sig{{s}}
\def\olig{{oligonucleotide}}  \def\oligs{{\olig s}}
\def\oligm{{oligomer}}  \def\oligms{{\oligm s}}
\def\kmer{{$k$-mer}} \def\kmers{{$k$-mers}}
\def\dist{{distribution}}  \def\dists{{\dist s}}
\def\kdist{{$k$-distribution}}  \def\kdists{{\kdist s}}
\def\lrt{{L_{r}}}
\def\sigs{{$s_u$}}   \def\sigl{{$s_L$}}  \def\sig{{s}} 
\def\ets{{$\eta_u$}}   \def\etl{{$\eta_L$}} 
\def\dist{{distribution}}  \def\dists{{\dist s}}
\def\foc{{occurrence frequency}}  \def\focs{{occurrence frequencies}}
\def\rmd{{root-mean-deviation}}
\def\mean#1{{\bar{#1}}}
\def\std{{std}}  \def\stds{{stds}}
\def\mset{{$m$-set}} \def\msets{{$m$-sets}}  
\def\MM{{\cal M}}  \def\MMs{{\MM_{s}}} \def\MMi{{\MM_R}}
\def\SS{{\cal S}} \def\QQ{{\cal Q}} \def\PP{{\cal P}} \def\GG{{\cal G}}
\def\ie{{\it i.e.}}
\def\etal{{\it et al.}}
\def\nrep{{$n$-replica}}   \def\nreps{{$n$-replicas}}   
\def\qr{{quasireplica}} \def\qrs{{quasireplicas}} \def\Qrs{{Quasireplicas}}
\def\qrn{{quasireplication}}
\def\uros{{underrepresented \oligs}}
\def\oros{{overrepresented \oligs}}
\def\kspec{{$k$-spectrum}}   \def\kspecs{{$k$-spectra}}
\def\kband{{$k$-band}}   \def\kbands{{$k$-bands}}
\def\lbar{{\bar l}}
\def\Bullet{{\large{$\bullet$}}}
\def\sqbull{{\vrule height 1.1ex width 1.0ex depth -.1ex }} 
\def\Circ{{\large{$\circ$}}}
\def\ApT{{(A+T)}}  \def\CpG{{(C+G)}}
\def\refeq#1{{Eq.~(\ref{#1})}}
\def\reffg#1{{Fig.~\ref{#1}}} \def\reffgs#1{{Figs.~\ref{#1}}}
\def\reftb#1{{Table~\ref{#1}}}  \def\reftbs#1{{Tables~\ref{#1}}}
\def\bgeq{\begin{equation}} \def\edeq{\end{equation}}
\def\bgeqy{\begin{eqnarray}} \def\edeqy{\end{eqnarray}}
\def\FFk{{\FF^{\{k\}}}} \def\FFkm{{\FF_m^{\{k\}}}}
\def\peqhalf{{$p$$\approx$0.5}} \def\pnehalf{{$p$$\ne$0.5}}
\def\Leff{{$L_{e}$}}  \def\Leffk{{$L_{e}(k)$}}   
\def\leff{{$l_{e}$}}  \def\leffk{{$l_{e}(k)$}}
\def\CVsq{{$CV^2$}}
\def\twocol#1#2{{\left(\begin{matrix}#1\cr #2\cr\end{matrix}\right)}}
\def\tworow#1#2{{\left(#1,#2\right)}}
\def\CVran{{CV^{\{ran\}}}} \def\sigran{{\sigma^{\{ran\}}}}
\def\CVlimit{{CV^{\{_{\infty}\}}}} 
\def\CVfluc{{CV_{fluc}}} \def\CVpfluc{{CV'_{fluc}}} 
\def\CVflucran{{CV_{fluc}^{\{ran\}}}}
\def\CVnf{{CV_{nflc}}}  
\def\CVorder{{CV_{order}}} \def\CVnfran{{CV_{nflc}^{\{ran\}}}}
\def\CVm{{CV_{m}}}
\def\signf{{\sigma_{nflc}}}  \def\sigfluc{{\sigma_{fluc}}}
\def\phig{{$\phi_g$}} \def\Iphi{{$I_\phi$}}
\newfont{\ssflarge}{cmssi17 scaled 1320}  
\newfont{\sfimed}{cmssi12 scaled 1200}    
\newfont{\sfi}{cmssi9 scaled 1200}        
\newfont{\sfb}{cmssdc10 scaled 1000}      
\newfont{\sfl}{cmssdc10 scaled 1320}      
\newfont{\nob}{cmb10 scaled 1200}         


 
\title
{\bf Genomes: at the edge of chaos with maximum information capacity}

\author{Sing-Guan Kong$^{1,2}$, Hong-Da Chen$^{1,2}$, 
Wen-Lang Fan$^{1,2}$, Jan Wigger$^4$, Andrew Torda$^4$, 
and H.C. Lee$^{2,3,5}$}

\affiliation{
$^1$Department of Physics, $^2$Graduate Institute of Biophysics, 
and $^3$Graduate Institute of Systems Biology and Bioinformatics, 
National Central University, Chungli, Taiwan 32001, ROC\\ 
$^4$Center for Bioinformatics, University of Hamburg, Hamburg, Germany\\
$^5$National Center for Theoretical Sciences, Hsinchu, Taiwan 30043, ROC
} 

\date{\today}


\begin{abstract}
We propose an order index, $\phi$, which quantifies the notion
of ``life at the edge of chaos'' when applied to genome
sequences. It maps genomes to a number from 0 (random and of 
infinite length) to 1
(fully ordered) and applies regardless of sequence length.
The 786 complete genomic sequences in GenBank 
were found to have  $\phi$ values in 
a very narrow range, 0.037$\pm$0.027.  
We show this implies that genomes are halfway towards 
being completely random, namely, at the edge of chaos.  
We argue that this narrow range represents the neighborhood of 
a fixed-point in the space of sequences, and genomes are 
driven there by the dynamics of 
a robust, predominantly neutral evolution process.
\end{abstract} 
\maketitle


The \emph{Edge of chaos} originally refers to the state of  
a computational system, such as cellular automata, when it is close to a 
transition to chaos, and gains the ability for complex information 
processing \cite{Langton90,Crutchfield90,Mitchell93}. 
The notion has since been used to describe biological states, and
life in general, on the assumption that life necessarily
involves complex computation \cite{Kauffman94}.  
In model systems such as cellular automata, there are well
defined procedures for recognizing the change in computational
cab ability during the transition from non-chaotic to chaotic
states \cite{Langton90,Mitchell93}.
However, these have not been adapted to the wider biological
context, even for the simplest of organisms.
But if we represent a living organism by its genome, view  
evolution as a dynamical process that drives genomes in the space  
of sequences, and consider chaos as a state of genome randomness, 
then we have a framework within which the meaning of ``life 
occurs at the edge of chaos'' may be investigated. 
Genomes, linear sequences written in 
the four chemical letters, or bases, A (adenine), C (cytosine), 
G (guanine) and T (thymine) and often referred to as books of life,  
regulate the functioning of organisms through the many kinds of 
codes embedded in them (there are also non-textual post-translational 
regulations; see, e.g. \cite{Davies00}).  
When genomes are seen as texts, they have several key properties
reflecting their complexity, including long-range correlations and scale
invariance \cite{Li92,Peng92,Voss92} (although this topic is
debated
\cite{Israeloff96}), self-similarity
\cite{Church93,Lu98,Nagai01,Chen05b}, and distinctive Shannon
redundancy \cite{Mantegna94,Stanley99,Chen05}.  However,
these properties do not give a measure of the proximity of a genome to
chaos or randomness.  Before the edge-of-chaos notion can be
explored, one needs to have a quantity that measures the
randomness of genomes as texts.   

Here we analyze genomes in terms of 
the frequency of occurrence of $k$-letter words, called \kmers, 
where $k$ is a small integer \cite{Hao00}. 
For a given $k$, the $4^k$ types of \kmers\ are partitioned into $k$+1 
``\msets'', $m$=0--$k$.  An \mset\ is composed of all the \kmers\ 
containing $m$ and only $m$ A or T's. There are 
$\tau_m$=$2^{k}${\footnotesize{$\twocol{k}{m}$}} types of \kmers\ in an \mset. 
The reason 
for partitioning the \kmers\ according to AT-content for statistical 
purposes is that
although the A:T and C:G ratios are invariably close to 1,
\cite{Prabhu93,Mrazek98,Bell99}, the AT to GC ratio may
differ significantly.  This partition is needed for preventing 
biased base composition from masking crucial statistical 
information in genomes \cite{Israeloff96,Chen05}. 
For $k$$\ge$2, the 
$k^{th}$ {\it order index} for a sequence of length $L$ (in bases) is
\bgeq
\phi \equiv {1\over (2-2(p^k +q^k))} \sum_m  {1\over L}
\left|L_m - L^{\{\infty\}}_m\right|, 
\label{e:order_index}
\edeq
where 0$<$$p$$<$1 is the fractional AT-content in 
the sequence; $q$=1-$p$; $L_m$ is the total number of \kmers\ 
in the \mset; and $L^{\{\infty\}}_m$ is 
the expected value for 
$L_m$ in a $p$-valued random sequence of infinite length:
$L^{\{\infty\}}_m$=$L$$2^{-k}$ $\tau_m$$p^m$$q$$^{k-m}$. 
The definition of $\phi$ is based on the observation that 
distribution-averages 
are useful indicators of the randomness of a sequence. 
The denominator on the right-hand-side of \refeq{e:order_index} 
is a normalization factor which ensures 
$\phi$$\approx$1 for an ordered sequence (in which all AT's are on,  
say, the 5' end and all CG's are on the 3' end).  
The singularities at $p$= 0 and 1 are not a practical problem
since no genome has such extreme base composition. 

From the central limit theory  
we expect, for random sequences, 
$|L_m -L^{\{\infty\}}_m|$ 
to scale as $L_m^{-1/2}$. We therefore expect $\phi$  to be proportional 
to $L^{-1/2}$ on average.  
\begin{figure}[h] 
\begin{center}
\vspace{0cm}
\includegraphics[width=3.0in]
{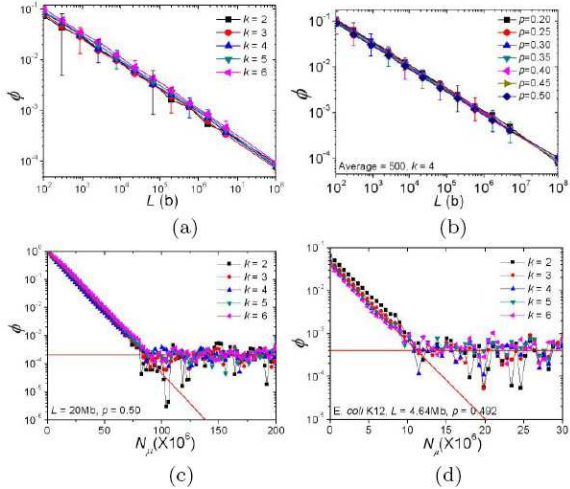}
\vspace{-10pt}
\caption{\label{f:randomOI-L} \footnotesize\sf{
(a) Log-log plot of order index, $\phi$, {\it vs.} length of random 
sequence for $p$=0.5 and $k$=2--6. (b) Same as (a); for $k$=4 and 
$p$=0.20--0.50.  
(c) Semi-log plot of $\phi$ {\it vs.} $N_\mu$, number of 
random point mutations, 
for an initially ordered 20 Mb, $p$=0.5 sequence. The 
intersection of the red lines 
is the critical point where sequence becomes random. 
(d) Same as (c); initial sequence is genome of \ecol. 
}}
\vspace{-0.8cm}
\end{center}
\end{figure}
The log-log plots in \reffg{f:randomOI-L} (a) and (b) show 
$\phi$ as a function of sequence length for different $k$'s and $p$'s. 
Each datum is averaged over 500 random sequences. It is seen that 
$\phi$ scales very well as $L^{-1/2}$ (with sizable fluctuations), 
and is only weakly dependent on $k$ and $p$. 
These results can be summarized for all $k$ and $p$ by an
empirical relation: 
\bgeq
\phi^{\{ran\}}=c_\phi L^{-\gamma_\phi}
\label{e:phi_ran}
\edeq 
with $\gamma_\phi=0.50\pm0.01$ and $c_\phi=1.0\pm0.2$ 
or, to a good approximation,
$\phi^{\{ran\}}$$\approx$$L^{-1/2}$.  This leads to the
convenient concept of an {\it equivalent length} for a $\phi$-value sequence, 
$L_{eq}(\phi)$$\equiv$$\phi^{-2}$, the nominal 
length of a random sequence whose order index is $\phi$.    

Random events such as point mutations acting on a non-random 
sequence decreases its 
order, and hence its $\phi$. 
\reffg{f:randomOI-L} (c)  shows that the $\phi$ of 
a $p$=0.5, 20 Mb ordered sequence,  decreases 
exponentially with the number of mutations $N_\mu$, 
until $N_\mu$ reaches a critical number $N_{\mu c}$. 
The critical value reflects the fact that a random sequence
does not become more random with further changes.
In other words, if one thinks of random point mutation as a dynamical  
action taking a sequence from one point in the sequence space to 
another, then a randomized sequence is a fixed-point of the action. 
Our studies of initially ordered sequences having a variety of lengths 
and base compositions yield, 
\bgeq
\phi = {\left\{\begin{matrix}
\exp{(-2N_\mu/L)},\ N_\mu \lesssim N_{\mu c};\cr
\phi_c\approx L^{-1/2},\quad N_\mu > N_{\mu c}\cr
\end{matrix}
\right.}
\label{e:phi_vs_mut}
\edeq
where the $N_{\mu c}$$\approx$(1/4)$L\ln L$, and the critical 
mutation rate is $\mu_c$$\equiv$$N_{\mu c}/L$$\approx$(1/4)$\ln L$. 
The formula for $N_{\mu c}$ compares well with simulation. In the case 
of \reffg{f:randomOI-L} (c), the coordinates of the 
simulation ($k$=4) critical point 
are ($\phi_c$, $N_{\mu c}$)=(2.2$\times$$10^{-4}$, 8.5$\times$$10^7$), 
as compared to the ``theoretical'' values 
(2.2$\times$$10^{-4}$, 8.4$\times$$10^7$). 
For typical sequences of genomic length ($L$$\sim$10$^{1\pm 1}$ Mb),   
$\mu_c$=4.0$\pm$0.6 mutations per base (b$^{-1}$). 
We use \refeq{e:phi_vs_mut} to assign to a $\phi$-valued sequence an 
{\it equivalent mutation rate}, 
$\mu_{eq}(\phi)$$\equiv$$\ln\phi^{-1/2}$, 
the nominal number of random point mutations per 
base required to bring the index of an ordered sequence to $\phi$.

\refeq{e:phi_vs_mut} can be adapted for application to sequences not 
initially ordered.  For example, the equivalent mutation rate 
for the 4.6 Mb genome of \ecol\ ($\phi$=0.049) is 1.5 b$^{-1}$.  
Since for a 4.6 Mb sequence $\mu_c$=3.8 b$^{-1}$, one expects  
an additional 2.3$\times$4.6$\times$10$^6$= 1.1$\times$10$^7$ 
mutations are needed to randomize it. 
In the simulation shown in \reffg{f:randomOI-L} (d),  
the actual number needed is found to be (1.1$\pm$0.1)$\times$10$^7$.  

We computed $\phi$ for 384 complete 
prokaryotic genomes (28 archaebacteria and 356 eubacteria) 
and 402 complete chromosomes from 28 eukaryotes  
of lengths ranging from 200 kb to 230 Mb.  
The rice genome was downloaded from the Rice Annotation Project Database 
\cite{rice}, and  all other sequences from 
the National Center for Biotechnology Information genome 
database \cite{NCBI}, during the period 26 Feb.--27 Nov., 2006.
The 28 eukaryotes (number of chromosomes and genome length 
in parenthesis) include 11 fungi, 
{\it A. fumigatus} (8, 28.8 Mb), {\it C. albicans} (1, 0.95 Mb),
{\it C. glabrata} (13, 12.3 Mb), {\it C. neoformans} (14, 19.1 Mb),
{\it D. hansenii} (7, 12.2 Mb), {\it E. cuniculi} (11, 2.50 Mb),
{\it E. gossypii} (7, 8.74 Mb), {\it K. lactis} (6, 10.7 Mb),
{\it S. cerevisiae} (Yeast) (16, 12.1 Mb), 
{\it S. pombe} (Fission Yeast) (3, 10.0 Mb), {\it Y. lipolytica} (6, 20.5 Mb);
the unicellular {\it P. falciparum} (Malaria) (14, 22.9 Mb);
2 plants, 
{\it A. thaliana} (Mustard) (5, 119 Mb), {\it O. sativa} (Rice) (12, 372 Mb);
5 insects, {\it C. elegans} (Worm) (6, 100 Mb), 
{\it D. melanogaster} (Fly) (6, 118 Mb),
{\it A. gambiae} (Mosquito) (5, 223 Mb), {\it A. mellifera} (Bee) (16, 183 Mb),
{\it T. castaneum} (Beetle) (10, 112 Mb);
9 vertebrates, {\it D. rerio} (Zebrafish) (25, 1.04 Gb), 
{\it G. gallus} (Chicken) (30, 933 Mb), {\it B. taurus} (Cow) (30, 1.41 Gb),
{\it C. familiaris} (Dog) (39, 2.31 Gb), {\it M. musculus} (Mouse) (21, 2.57 Gb),
{\it R. norvegicus} (Rat) (21, 2.50 Gb),
{\it M. mulatta} (Monkey) (21, 2.73 Gb),
{\it P. troglodytes} (Chimpanzee) (25, 2.86 Gb),
{\it H. sapiens} (Human) (24, 2.87 Gb).

\begin{figure}[h] 
\begin{center}
\includegraphics[width=3.15in]
{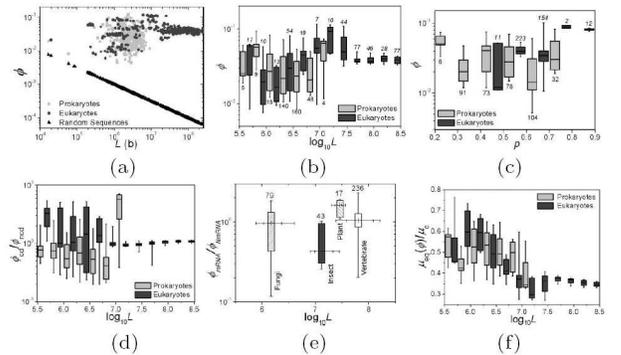}
\vspace{-8pt}
\caption{\label{f:IO-genome_histogram} \footnotesize\sf{
(a) Order index, $\phi$, {\it vs.} sequence length, $L$,
for 384 prokaryotic genomes (gray $\sqbull$'s), 
402 eukaryotic chromosomes (black \Bullet's), and random 
sequences (line composed of $\bigtriangledown$'s). 
In (b--f):  Box (gray for prokaryotes; black for eukaryotes) 
height is given by 25\% to 50\% values 
and the range represents 10\% to 90\% values; numbers above boxes are
numbers of sequences in group; all $\phi$'s are averaged over $k$=2 to 6.
(b) $\phi$ {\it vs.} $\log L$.  
(c) $\phi$ {\it vs.} fractional AT-content, $p$.
(d) Ratio of $\phi_{cd}$ (for coding parts) 
to $\phi_{ncd}$ (noncoding) {\it vs.} $\log L$.
(e) Ratio of $\phi_{mRNA}$  (mRNA segments) 
to $\phi_{nmRNA}$ (non-mRNA), averaged over classes of eukaryotes.  
(f) Ratio of equivalent mutation rate, $\mu_{eq}$, to critical 
mutation rate, $\mu_c$, {\it vs.} $\log L$.}} 
\vspace{-0.5cm}
\end{center}
\end{figure}

The results shown in \reffg{f:IO-genome_histogram} 
indicate that genomic $\phi$'s 
systematically vary neither with sequence length 
((a) and (b)), 
nor with base composition ((c)). 
Instead they have a nearly universal value --- the  
average over all sequences is \phig$\equiv$0.037$\pm$0.027 
(this defines the symbol \phig). 
We have verified that, as a general rule, within a genome 
the variation in segmental $\phi$ decreases with segmental length 
and the average $\phi$ reaches its 
whole-genome value when the size of the segment exceeds 50 kb. 
In \reffg{f:IO-genome_histogram} (a) the spread in $\phi$ of the genomic 
data shows a tendency to decrease with sequence length. Part of this effect  
may be purely statistical: smaller sample sizes ({\it i.e.}, sequence lengths) 
tend to have larger statistical fluctuations. Part of it may also be 
because sequences 
longer than 10 Mb are all from chromosomes of multicellular eukaryotes 
that are phylogenetically close. 
In any case 
\reffg{f:IO-genome_histogram} (a) clearly puts the genomes in a category 
apart from random sequences. 

From each complete sequence, we extracted the coding and noncoding parts 
(owing to imperfect annotation, the sum of the parts sometimes differ 
slightly from the whole), 
then concatenated the parts into two separate sequences and 
computed their order indexes, $\phi_{cd}$ and $\phi_{ncd}$, respectively.  
A summary of the ratio $\phi_{cd}/\phi_{ncd}$ for sets of genomes grouped 
by length is given in  \reffg{f:IO-genome_histogram} (d).   
For prokaryotes the  ratio ranges (10$^{th}$ to 90$^{th}$ percentile) 
from  0.15 to 3 with a median of about 0.5.
Notable exceptions are the three bacteria 
with exceptionally large genomes ($L$$\lesssim$10 Mb) with 
ratios ranging from 5 to 7: 
\save, \scoe, and {\it Mycobacterium sp. MCS}. 
 
For the eukaryotic chromosomes longer than 10 Mb  
the ratios do not significantly deviate from unity.  
Mustard, whose coding and 
noncoding parts have nearly equal lengths ($\sim$10--12 Mb), 
is the only exception in this category 
with $\phi_{cd}/\phi_{ncd}$$\approx$7 
(these ratios are beyond the 90 percentile 
and therefore are not included in \reffg{f:IO-genome_histogram} (d)). 
In this case $\phi_{cd}$$\approx$0.055 is similar to other genomes 
while $\phi_{ncd}$$\approx$0.0075 is about seven times less than the norm.  
Rice, the only other plant 
included in this study, with $\phi_{cd}/\phi_{ncd}$$\approx$0.35 
is unlike mustard but more like the other eukaryotes. 
For the eukaryotic chromosomes shorter than 10 Mb  
the ratios average to about 2 but show greater variation.   

The coding parts of eukaryotic genomes are further partitioned into 
mRNA and non-mRNA parts, and their $\phi$'s computed separately. 
Averaged over sets of 
organisms, $\phi_{mRNA}/\phi_{nmRNA}$ is of the order of 1, 
with the ratio being $\sim$0.5 for insects and $\sim$2 for plants
(\reffg{f:IO-genome_histogram} (e)). 
For the latter, the ratio is $\sim$1 
for the five chromosomes of mustard and $\sim$2 for the twelve 
chromosomes of rice.  
In summary, the differences in $\phi$ between coding and 
non-coding parts, and between mRNA and non-mRNA parts are much smaller  
than the difference between genomes and random sequences.  

The ratio $\mu_{eq}(\phi)$/$\mu_c$ is an indication of how close  
a sequence is to being random.
\reffg{f:IO-genome_histogram} (f) shows that 
the shorter ($L$$\gtrsim$10 Mb) sequences 
are roughly half-way,  and the longer sequences, 
one-third of the way, towards becoming random.   
The systematic but weak length-dependence of the ratio is 
explained by the fact that the genomic $\phi$, hence $\mu_{eq}(\phi)$, 
is approximately constant, whereas $\mu_c$ is proportional to $\ln L$.  
The overall average of the ratio is 0.45$\pm$0.11.

\begin{figure}[h] 
\vbox{
\begin{center}
\vspace{0cm}
\hspace{10pt}\includegraphics[height=2.2in]
{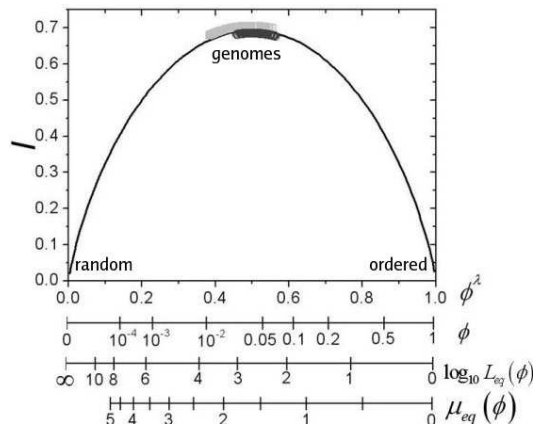}
\caption{\label{f:I} \footnotesize\sf{
The function $I(z)$ (\refeq{e:info_capacity}) 
plotted as a function of: $z$=$\phi^\lambda$ 
($\lambda$=0.21);  $\phi$; $\log_{10}L_{eq}(\phi)$ (\refeq{e:phi_ran});  
$\mu_{eq}(\phi)$ (in units of b$^{-1}$; \refeq{e:phi_vs_mut}). 
Data from prokaryotic (gray) and eukaryotic (black) genomes 
occur near the peak of $I$ and have $\phi$$\sim$\phig,
$L_{eq}$$\sim$.25--10 kb, and $\mu_{eq}$$\sim$1.8$\pm$0.5 b$^{-1}$.}} 
\end{center}
} 
\vspace{-0.5cm}
\end{figure}

We summarize our results by considering the function 
\bgeq
I(z) =  -z \ln z -(1- z)\ln(1- z) 
\label{e:info_capacity}
\edeq
where $z$=$\phi^{\lambda}$ and $\lambda$=0.21.  
The value of the exponent $\lambda$ is determined by requiring that 
$z$=0.5 at $\phi$=\phig.  $I(z)$ is the simplest function that
maps the range (0,1) to a positive real value,
has zeros at (and only at) $z$=0  and 1,
has a maximum at $z$=0.5 and  
is symmetric with respect to the point $z$=0.5.  
In \reffg{f:I} the parabola-like curve shows  
$I(z)$ plotted against $z$. 
In addition, three other sets of abscissas are given: $\phi$; 
$\log_{10}L_{eq}(\phi)$, where $L_{eq}(\phi)$ is the equivalent length 
(\refeq{e:phi_ran}); and $\mu_{eq}(\phi)$, the equivalent mutation rate. 
A scale linear in $z$, relative to one in $\phi$, is a better 
representation of the space of possible sequence lengths. 
It is seen in \reffg{f:I} that genomes are 
concentrated near the peak of the $I$-curve and equally and far removed 
from the random ($z$$\sim$0) and ordered ($z$$\sim$1) sequences. 

The genomic equivalent lengths, occupying a small neighborhood 
around at $L_{eq}(\phi_g)$=730 b, are far shorter than the actual 
lengths of complete sequences.  Among the many possible mechanisms that may 
cause long sequences to have short equivalent lengths, by far the  
simplest is replication. This is because  
a long sequence of length $L$ composed of multiple replications of a random 
sequence $l$ bases long will have $L_{eq}$$\sim$$l$, independent of $L$.  
Similarly, if genome growth is dominated by random segmental duplication
 \cite{Lynch02,Bailey02,Zhang05}, 
then the genomic $L_{eq}$ will be much shorter genome length \cite{Chen05}.

The genomic equivalent mutation rates span a small range around 
1.8 b$^{-1}$, or about 45\% of the 
critical mutation rate  
of approximately 4 b$^{-1}$ that would randomize the genomes.  
Thus, for example, a typical worm (\cele) 
chromosome, with an average length of 17 Mb and an equivalent mutation 
rate of 1.8 b$^{-1}$,  is as random 
as an initially ordered 17 Mb sequence after having undergone    
31 million random mutations - as compared to the 
68 million mutations which would randomize the sequence. 
In this sense genomes are quasi-random - or ``at the edge of chaos''.  
For a linear text, quasi-randomness satisfies two crucial necessary 
conditions for high information content: high efficiency 
and large variation in word usage.  A random sequence has maximum 
word-usage efficiency because all its \kmers\ 
in an \mset\ have occurrence frequencies 
very close to the theoretical mean frequency of the set, 
$\bar f^{\{\infty\}}_m$=$L^{\{\infty\}}_m$/$\tau_m$.  
However, this also implies minimum word-usage 
variation, which prevents a random sequence from being information-rich.  
In a quasi-random sequence a compromise between high efficiency and 
large variation in word usage is obtained 
by suitably relaxing the equal-frequency condition \cite{Chen05}, thus 
allowing a genome at the edge of chaos to have close to 
maximum information {\it capacity}.  
 
The high concentration of genomic $\phi$'s near \phig\  
may be interpreted as the signature of a certain robust characteristics  
in the genomic evolution processes. The near equality of 
$\phi$'s for coding and noncoding regions within a genome 
suggests that the underlying evolution processes are 
not dominated by codon selection, but are likely 
predominantly selectively neutral  \cite{Kimura80,Fu93}. 
We therefore propose the following conjecture: 
Just as randomness is a fixed-point 
of the action of random point mutations, the state of genomes 
defined by $\phi$$\sim$\phig\ is a fixed-point of the action of a 
robust, predominantly neutral evolution process.  
The observed shortness of $L_{eq}(\phi_g)$ suggests 
that the neutral process is dominated by (non-deleterious) random segmental 
duplications \cite{Lynch02,Bailey02,Zhang05}, 
occurring singly  \cite{Chen05,Hsieh03} and in tandem \cite{Messer05}. 
We consider random segmental duplication to be an infrastructure-building 
process because it does not necessarily produce information directly.  
Instead,  it causes genomic $\phi$ to be close to \phig, 
giving genomes maximum information capacity.  
Since this enhances genomic fitness indirectly, the neutral process  
may in itself be a product of natural selection. 
The near randomness of the neutral process guarantees the fixed-point 
associated with \phig\ to have a very large configuration space, 
hence relatively low free energy, thus rendering \phig-valued states 
widely accessible.   
In contrast, non-neutral, information-gathering processes dominated by  
selection (narrowly construed) are predominantly point mutations:   
they are poor mechanisms for inducing genomic states of maximum capacity, 
and do not lead to widely accessible states.    
Taken together these suggest that the evolution of 
the genome may have been driven by a two-stage process: 
one neutral, robust, infrastructure-building and universal,  
and the other selective, fine-tuning, information-gathering and diverse. 
An example of such a two-step process is found in the paradigm 
of accidental gene duplication followed by mutation driven 
subfunctionalization \cite{Lynch00,Zhang03}. 
We may assume that during the long history of the genome's growth and
evolution, the twin-processes acted in a ratchet-like, complementary
manner, driving the genome, in successive stages, to a state of
maximum information capacity, and helping it to acquire, at 
each stage, near-maximum information content.

This work is supported in part by grant nos. 95-2311-B-008-001 
and 95-2911-I-008-004 from the National Science Council (ROC).  





\begin{thebibliography}{99}
\bibitem{Langton90}
C. G. Langton. 
{\it Physica D} {\bf  42}, 12-37 (1990). 
\bibitem{Crutchfield90}
J. P. Crutchfield and K. Young. 
In W. H. Zurek, editor, {\it Complexity, entropy, and the physics of 
information}, 223-269 (Addison-Wesley, Redwood City, CA, 1990).
\bibitem{Mitchell93}
M. Mitchell, P. T. Hraber, and J. P. Crutchfeld. 
{\it Complex Systems}, {\bf 7}, 89-130 (1993).
\bibitem{Kauffman94}
S. A. Kauffman. 
(Oxford Univ. Press, London, 1993).
\bibitem{Davies00}
S. P. Davies,  \etal\ 
{\it Biochem. J.} {\it 351}, 95-105 (2000).
\bibitem{Li92} 
W. Li and K. Kaneko. 
{\it Europhys. Lett.} {\bf 17}, 655-660 (1992). 
\bibitem{Peng92} 
C. K. Peng, \etal\  
{\it Nature} {\bf 356}, 168-170 (1992); 
{\it Phys. Rev. E} {\bf 47}, 3730-3733 (1993).
\bibitem{Voss92} 
R. F. Voss. 
{\it Phys. Rev. Lett.} {\bf 68}, 3805-3808 (1992).
\bibitem{Israeloff96} 
N. E. Israeloff, \etal\  
{\it Phys. Rev. Lett.} {\bf 76}, p1976 (1996);
S. Bonhoeffer, \etal\  
{\it loc. cit.} p1977; 
R. F. Voss. 
{\it loc. cit.} p1978;  
N. Mantegna, \etal\  
{\it loc. cit.} p1979. 
Also: 
C. A. Chatzidimitriou-Dreismann,  \etal\ 
{\it NAR} {\bf 24}, 1676-1682 (1996).  
\bibitem{Church93}
K. W. Church, J. I. Helfman, {\it J. Comp. Graph. Stat.} {\bf 2}, 153-174 (1993). 
\bibitem{Lu98} 
X. Lu, Z. Sun, H. Chen, Y. Li. 
{\it Phys. Rev. E} {\bf 58}, 3578-3584 (1998).
\bibitem{Nagai01} 
N. Nagai. 
{\it Jpn J Physiol.} {\bf 51}, 159-68 (2001).
\bibitem{Chen05b} 
T. Y. Chen, L. C. Hsieh and H. C. Lee.
{\it Comp. Phys. Comm.} {\bf 169}, 218-221 (2005).
\bibitem{Mantegna94} 
R..N. Mantegna, \etal\ 
{\it Phys. Rev. Lett.} {\bf 73}, 3169-3172 (1994); 
{\it Phys. Rev. E} {\bf 52}, 2939-2950 (1995).  
\bibitem{Stanley99} 
H. E. Stanley, \etal\  
{\it Physica A} {\bf 273}, 1-18 (1999). 
\bibitem{Chen05} 
H. D. Chen, \etal\  
{\it Phys. Rev. Lett.} {\bf 94} 178103 (2005).
\bibitem{Hao00}
B. L. Hao, H. C. Lee and S. Y. Zhang.
{\it Chaos Solitons Fract.} {\bf 11}, 825-836 (2000).
\bibitem{Prabhu93}
V. V. Prabhu
{\it Nucl. Acids Res.} {\bf 21}, 2797-2800 (1993).
\bibitem{Mrazek98}
J. Mrazek and S. Karlin.
{\it Proc. Nat. Acad. Sci.} (USA) {\bf 95} 3720-3725 (1998).
\bibitem{Bell99}
S. J. Bell and D. R. Forsdyke
{\it J. theor. Biol.} {\bf 197}, 51-61 (1999). 
\bibitem{rice} Rice Annotation Project Database; 
http://rapdb.lab.nig. ac.jp/ (Nov. 2006).
\bibitem{NCBI}  
National Center for Biotechnology Information genome 
database; http://www.ncbi.nlm.nih.gov/ (Oct. 2006).  
\bibitem{Lynch02}
M. Lynch. 
{\it Science} {\bf 297} 945-947 (2002).
\bibitem{Bailey02}
J. A. Bailey, \etal\ 
{\it Science} {\bf 297}, 1003-1007 (2002).
\bibitem{Zhang05}
L. Zhang, \etal\ 
{\it Mol. Bio. Evol.} {\bf 22} 135-141 (2005).
\bibitem{Kimura80}  
M. Kimura. 
{\it J. Mol. Evol.} {\bf 16} 111-120 (1980). 
\bibitem{Fu93}
Y. X. Fu and W. H. Li. 
{\it Genetics} {\bf 133} 693-709 ( 1993).
\bibitem{Hsieh03}
L. S. Hsieh, L. F. Luo, F. M. Ji and H. C. Lee. 
{\it Phys. Rev. Lett.} {\bf 90} 018101 (2003).
\bibitem{Messer05}
P. W. Messer, P. F. Arndt, and M. Laessig.
{\it Phys. Rev. Lett.} {\bf 94} 138103 (2005). 
\bibitem{Lynch00}
M. Lynch and J. S. Conery.
{\it Science} {\bf 290} 1151-1155 (2000).
\bibitem{Zhang03}
J. Zhang. 
{\it Trends Eco. Evol.} {\bf 18}, 292-298 (2003). 
\end{thebibliography}
\end{document}